\documentstyle[titlepage]{article}
\begin{document}
\title{Magnetic fields and rotation of spiral galaxies}
\author{E. Battaner, H. Lesch and E. Florido} 
%\address{Dpto. Fisica Te\'orica y del Cosmos. Universidad de Granada. Avda. Fu%entenueva s/n. 18071 Granada. Spain and Institut f\"{u}r Astronomie und Astrop%hisik der Universit\"{a}t M\"{u}nchen. Scheunerstrasse 1, D-81679. M\"{u}nchen%. Germany} 
\maketitle
\begin{abstract} 
We present a simplified model in which we suggest that two important galactic problems -the magnetic field configuration at large scales and the flat rotation curve- may be simultaneously explained. A highly convective disc produces a high turbulent magnetic diffusion in the vertical direction, stablishing a merging of extragalactic and galactic magnetic fields. The outer disc may then adquire a magnetic energy gradient very close to the gradient required to explain the rotation curve, without the hypothesis of galactic dark matter. Our model predicts symmetries of the galactic field in noticeable agreement with the large scale structure of our galaxy. 
\end{abstract}

\section{Introduction}
Magnetic fields have been proposed as the main cause of the fast rotation in the outer disc of spiral galaxies (Nelson, 1988; Battaner et al. 1992; Battaner and Florido, 1995) thus providing an alternative explanation to the well-known hypothesis of dark matter within the galaxy. It was shown in these papers that an assumed magnetic field distribution can reproduce the rotation curve. In this short note, we undertake the complementary question: what mechanisms are taking place in the disc to produce those magnetic fields which are precisely required to explain the rotation curve.

The question of the origin and present distribution of galactic magnetic fields is still under debate. The classical dynamo theory (e.g. Wielebinski $\&$ Krause, 1993) cannot offer a clear scenario.
The standard
dynamo approach was critizized for several reasons: First the back reaction of the turbulence on the amplified magnetic field may be very strong on small scales, preventing a further amplification on large-scales, as necessary for galactic magnetic fields (Kulsrud 1986; Kulsrud and Anderson 1992). This critics concluds that galactic magnetic fields must be of primordial origin, i.e. the fields originate from processes before galaxies form.

The second critics concerns the appropriate velocity field. In standard dynamo theory an axisymmetric velocity field is used, which mainly describes the effects of differential rotation. Common nonaxisymmetric features like spiral arms, bar and tidal interactions are completely neglected in this approach. An alternative theory includes this structures and comes to the result that they can have an important influence on the structure and strength of the field (Chiba and Lesch 1994; Lesch and Chiba 1997).

Since the dynamo scenario is so unclear we follow here the scenario that galaxies form in an already magnetized intergalactic medium with a field of about $1\, \mu$G, as indicated by the following observations (Kronberg 1994): 

Kim et al. (1990) estimated 1.7 $\pm$0.9 $\mu$G in the Coma cluster, core region and Feretti et al (1995), 6$\mu$G. Deiss et al (1997) found values this order of magnitude, 3 Mpc away from the center of Coma. Similar values were obtained by Vallee et al (1986) in A 2319. Using a sample of 53 Abell clusters, Kim et al (1991) obtained typical values of $\sim$1$\mu$G extending typically to $\sim$0.5 Mpc from the centres. In "cooling-flow" galaxy clusters with cD galaxies in the centre like Cygnus A (Dreher et al. 87), M87 (Owen et al. 1990), Hydra A (Taylor et al. 1990), 3C295 (Perley and Taylor 1991) and Abell 1795 (Ge and Owen) large-scale magnetic fields have been detected via Faraday rotation measures up to a scale
of 100 kpc (Taylor and Perley (1993). The field strength are at least several $\mu$G, even in 3C295 at a redshift of 0.461 such field strengths have been measured (Perley and Taylor 1991).

These results suggest {\it prima facie} that at clusters individual galaxies may form out of a strongly magnetized environment having prior magnetic field strengths at least as large as in the interstellar medium of nearby galaxies like M81 or M31. 

There exist a lower limit of 0.3-0.5 $\mu$G implied by the absence of inverse-Compton-generated X-rays (e.g. Gursky $\&$ Schwartz, 1977; Raphely $\&$ Gruber, 1988). Garrington et al (1988) and Garrington and Conway (1991) also obtained a typical value of 1-2 $\mu$G, suggesting that $\mu$G-level magnetic fields are widespread (Kronberg, 1994). 

Also important is the observational evidence that galaxies are highly magnetized even at protogalactic stages, as detected in Absorption-Line-Systems in quasars. Kronberg $\&$ Perry (1982) and Watson $\&$ Perry (1991) estimated typical values of few microgauss. Observations of Lyman-$\alpha$ clouds at redshifts of z=2 show magnetic fields of the order of 3 $\mu$G (Wolfe et al. 1992). The mechanism of producing magnetic fields must therefore act very fast or galaxies were already formed with them. Similar values have been obtained for high redshifted galaxy discs (Wolfe, 1988; Kronberg et al 1992). Giving the similarity of all these values, an efficient connection between extragalactic and galactic fields have existed since the birth of spiral galaxies. This efficient connection will be here assumed to be turbulent magnetic diffusion. 

There exist in fact large turbulent motions connecting the disc and the corona and the intergalactic medium. Fountain models (Shapiro $\&$ Field, 1976; Breitschwerdt et al. 1991; Kahn, 1994) provide a very active convection. Other effects, such as Parker instabilities also contribute (Hanasz $\&$ Lesch, 1993). Sofue, Wakamatsu $\&$ Malin (1994) have identified in NGC 253 dark filaments and lanes, arcs and other microstructures illustrating not only complex and effective convection mechanisms, but even a magnetic field transport. Indeed, vertical velocity dispersions of $\sim$100 km/s are not uncommon in normal HII regions (Rozas, 1996). 

All this justifies the following scenario: The origin of disc magnetic fields is extragalactic. Extragalactic magnetic fields are large and can easily penetrate into the disc through the active convection. Within the disc, magnetic fields are slightly amplified and re-ordered by rotation.

With about 1 $\mu$G in the intergalactic medium and about 10$\mu$G in the inner disc, it is difficult to conceive the outer disc with less than 1$\mu$G. With this field strength and typical densities in this zone, Alfven velocities become of the order of 100 km $s^{-1}$, similar to the rotation velocity. It is then obvious that the action of magnetic fields in the outer discs cannot be neglected at all. It will be shown that they could produce the required centripetal force to give a fast rotation.

Probably other mechanisms are at work, rendering the final picture a much more complicated one. But we prefer to concentrate on this effect as it has not been explored. Merging of inter and extragalactic magnetic fields cannot be neglected at all and mechanisms other than dynamo are required. Magnetic fields must have a decisive influence in the dynamics of the disc.

\section{The simplified model}
In Battaner and Florido (1995) it was shown that the azimuthal magnetic field curve required to produce a centripetal force to explain the real rotation curves, could be written as \begin{equation}
B_\phi =B_{\phi}^* +B'_\phi
\end{equation}
where $B_{\phi}^* \propto r^{-1}$ is called the critical profile and $B'_\phi$ can be considered as a corrective term as $B'_\phi (r \rightarrow \infty)\rightarrow 0$. Then, a zero order step for the understanding of the stablishment of this azimuthal field would be the identification of a mechanism responsible for $B_{\phi}^*(r)$. 

Let us adopt the standard expression for the induction equation (e.g. Ruzmaikin et al, 1988)

\begin{equation}
{{\partial B_r}\over {\partial t}}=\beta {{\partial}\over {\partial z}} \left( {{\partial B_r}\over {\partial z}}- {{\partial B_z}\over {\partial r}} \right) \end{equation}
\begin{eqnarray}
{{\partial B_\phi} \over {\partial t}} = B_r {{\partial \theta} \over {\partial r}} + B_z {{\partial \theta} \over {\partial z}} - \theta {B_r \over r} + \nonumber \\
\beta \left( {{\partial^2 B_\phi} \over {\partial z^2}} + {\partial \over {\partial r}} \left( {{\partial B_\phi}\over {\partial r}} + {B_\phi \over r} \right) \right) \end{eqnarray}
\begin{equation}
{{\partial B_z} \over {\partial t}} =
- \beta \left( {{\partial} \over {\partial r}} \left( {{\partial B_r} \over {\partial z}}-{{\partial B_z}\over {\partial r}} \right) + {1 \over r} \left( {{\partial B_r} \over {\partial z}} - 
{{\partial B_z} \over {\partial r}} \right) \right) \end{equation}

where $B_r$, $B_\phi$ and $B_z$ are the radial, azimuthal and vertical components of the magnetic field strength, $\theta$ is the rotation velocity and $\beta$ the coefficient of turbulent magnetic diffusion. No $\alpha$-term representing any dynamo action has been included. Some amplification mechanism is probably not ignorable, but let us first consider the ideal case in which the boundary conditions suply as much magnetic fields as necessary. 

To estimate the coefficient $\beta$ let us use the standard expression \begin{equation}
\beta = {1 \over 3} v l \approx 2 \times 10^{27} cm^2 s^{-1} \approx 6 kpc^2 Gyr^{-1}
\end{equation}

To estimate the order of magnitude we may assume that $l$, the typical length of the disc-corona convective cells is about 1 kpc and $v$, the typical convection velocity, of the order of 20 km $s^{-1}$. This is an intermediate value between a typical value in the inner disc (e.g. $10^{26} cm^2 s^{-1}$ from Ruzmaikin and Sokoloff, 1988) and in the corona (e.g. $5 \times 10^{27} cm^2 s^{-1}$ from Sokoloff and Shukurov, 1990). Ruzmaikin, Sokoloff and Shukurov (1989) proposed $8 \times 10^{29} cm^2 s^{-1}$ for the intergalactic medium in a cluster. The characteristic diffusion time is $L^2/\beta$, which gives, for the characteristic length $L \approx 1$kpc, a typical value of 0.2 Gyr. Therefore, turbulent magnetic diffusion may be really a very fast process.

We note that the adopted value for $\beta$ is a conservative estimate since it is clear from observations that galaxies go through an initial starburst phase (Heckman 1997 and references therein). Such violent stellar activity drives strong magnetized outflows as can be observed in nearby starburst objects like M82 filling a halo of several kpc (Reuter et al. 1992). High velocity plasma streams injected into the halo on scales of kpc will definitely result in a considerably higher turbulent diffusion than the value we adopted. 

Not all components of the field penetrate with the same facility. We consider the extragalactic field varying in typical scale lengths much greater than the galactic diameter, and having an arbitrary direction. The component $B_z$ easily penetrates and magnetic diffusion acts so fast than we may assume $B_z$=constant in the whole outer disc, with a value equal to the extragalactic value. (We are only considering the outer disc and its neighborhood in the vertical direction).

For the radial component, the situation is different. For instances, if a given $B_r$-component penetrates into the disc at a given time and at point $(r, \phi)$, rotation would transport the frozen-in magnetic field lines into the opposite position ($r, \phi +\pi$) in half a rotation period. There, the direction of the field vector would be opposite to the $B_r$-component penetrating from outside, provided the homogeneity of extragalactic magnetic field. Reconnection would proceed, so as we may assume $B_r=0$ at the boundaries. 

The situation for $B_\phi$ at the boundaries is the same than for $B_r$, so we assume $B_\phi =0$ at the boundaries. But as far as $B_\phi$ is easily amplified by rotation and can be generated from $B_z$ which is not vanishing, we do not assume $B_\phi =0$ in all the considered region; rather $B_\phi(r)$ is what we should obtain. 

Let us also assume steady state condition, $\partial /\partial t=0$. With $B_z$=constant, equation (2) informs us that $\partial^2B_r/\partial z^2$=0. If the reconnection zone at the vertical boundaries is extended, we can assume not only $B_r (r \rightarrow \infty) \rightarrow 0$, but also $\partial B_r/\partial z (z \rightarrow \infty) \rightarrow 0$, therefore, integration of this equation simply gives $B_r =0$ in the whole space considered. This result is fully compatible with equation (3). We then have for equation (2)
\begin{equation}
0= B_z {{\partial \theta} \over {\partial z}} + \beta \left( {{\partial^2 B_\phi} \over {\partial z^2}} + {\partial \over {\partial r}} \left(
{{\partial B_\phi} \over {\partial r}} + {B_\phi \over r} \right) \right)
\end{equation}
This equation would provide the required solution $B_\phi (r,z)$ provided that $\theta (z)$ is known. As we are not interested in a precise integration, let us assume $\partial \theta /\partial z =0$. This is not unreasonable given the relative low thickness of the disk. As we are aware a clear variation of $\theta$ with $z$ has never been found.

We cannot assume $\partial^2 B_\phi /\partial z^2 =0$ as $B_\phi=0$ at the boundaries and non-vanishing in the plane. But it should be close to zero in a wide region close to the plane. The decrease of magnetic field strength with $z$ is observed to be very slow, not only in galaxies with a radio-halo (Beck et al 1994 for NGC 253; Dahlem et al. 1994, for NGC 891 and Dahlem et al.
1995, for many other galaxies) but even for spiral galaxies without a radio-halo (Ruzmaikin et al, 1988; Wielebinski, 1993) and for our own Galaxy (Han and Qiao, 1994). From dwarf starburst galaxies would be the extreme case of extended magnetized halos (Reuter et al, 1992) even able at early times to magnetize a large fraction of the intergalactic space (Kronberg and Lesch, 1997). Therefore, for $z$ small, near the plane we may consider negligible the variations of $B_\phi$ in the vertical direction, in order to annalyze the physical content of equation (2). For small $\mid z \mid$, then \begin{equation}
{\partial \over {\partial r}} \left(
{{\partial B_\phi} \over {\partial r}} + {B_\phi \over r} \right) =0
\end{equation}
or
\begin{equation}
B_\phi r =constant
\end{equation}
Hence
\begin{equation}
B_\phi \propto {1 \over r}
\end{equation}
which is precisely the critical profile. Therefore the effect of turbulent magnetic diffusion is to provide a magnetic energy gradient close to that required to explain rotation curves without the help of dark matter.

\section{Conclusions}

As there is increasing evidence that extragalactic magnetic fields are of the order of 1 $\mu$G, slightly lower than typical values inside the disc, and as there is increasing evidence that a very active convection connects the disc and the corona, future models should not neglect this effect. This connection ensures magnetic field strengths in the outer disc of about 1$\mu$G or larger. Therefore, future models should not neglect the influence of magnetic fields on the dynamics of the outer disc.

Here, we present a model, simplified and dealing with an idealized galaxy, in which we have taken these effects as dominant. The result suggests that the merging of both, galactic and extragalactic magnetic fields is supported by turbulent magnetic diffusion, and explains the fast rotation curve.

The simmetries predicted in our model of the magnetic field are in agreement with recent analyses of rotation measurements by Han et al. (1997) for the magnetic field in our own galaxy. In this work it is proposed that the so called AO mode prevails in our galaxy. Our model includes no dynamo action but the obtained large-scale structure has much in common with the AO mode. We also predict an antisymmetry of the azimuthal field in both hemispheres and the same direction of the vertical field in both hemispheres for $|l| < 90^\circ$. This AO dynamo mode has been also observed in other galaxies, but in view of the symmetry similarities with our predictions these galaxies can be interpreted by another different theoretical scenario: they may result from an effective interconection with intergalactic magnetic fields. 

\vskip 2cm
\noindent {\bf Bibliography}

\noindent Battaner, E. \& Florido, E. 1995, MNRAS, 277, 1129 

\noindent Battaner, E., Garrido, J.L., Membrado, M. \& Florido, E. 1992, Nature, 369, 652

\noindent Beck, R., Carilli, C.L., Holdaway, M.A. \& Klein, U. 1994, AA, 292, 409 

\noindent Breitschwerdt, D., McKenzie, J.F. \& V\"{o}lk, H.J. 1991, AA, 245, 79 

\noindent Chiba, M. \& Lesch, H. 1994, AA, 284, 731 

\noindent Dahlem, M., Dettmar, R.J. \& Hummel, E. 1994, AA, 290, 384 

\noindent Dahlem, M., Lisenfeld, U. \& Golla, G. 1995, ApJ, 444, 119 

\noindent Deiss, B.M., Reich, W., Lesch, H. \& Wielebinski, R. 1997, to be published in AA.

\noindent Dreher, J.W., Carilli, C.L., Perley, R.A., 1987, ApJ 316, 611 

\noindent Feretti, L., Dallacasa, D., Giovannini, G. \& Tagliani, A. 1995, AA, 302, 680

\noindent Garrington, S.T., Lehay, J.P., Conway, R.G. \& Laing, R.A. 1988, Nature, 331, 147

\noindent Garrington, S.T. \& Conway, R.G. 1991, MNRAS, 250, 198 

\noindent Ge, J.P., Owen, F.N., 1993, AJ 105, 778

\noindent Gursky, H. \& Schwartz, D.A. 1977, Ann. Rev. Astron. Astrophys., 15, 553 

\noindent Han, J.L. \& Qiao, G.J. 1994, AA, 288, 759 

\noindent Han, J. L., Manchester, R. N., Berkhuijsen, E. M. \& Beck, R. 1997, A\&A 322, 98

\noindent Hanasz, M. \& Lesch, H. 1993, AA, 278, 561 

\noindent Heckman, T.M., 1997, in {\it Star FormationNear and Far}, eds. S. Holt and L. Mundy, AIP, New York 

\noindent Kahn, F.D. 1994, Ap. J. S., 216, 325

\noindent Kim, K.T., Kronberg, P.P., Dewdney, P.E. \& Landecker, T.L. 1990, ApJ, 355, 29

\noindent Kim, K.T., Tribble, P.C. \& Kronberg, P.P. 1991, ApJ, 379, 80 

\noindent Kronberg, P.P. 1994, Reports on Progress in Physics, 57, 325 

\noindent Kronberg, P.P. \& Lesch, H. 1997, in Physics of the Galactic Halos, ed. by H. Lesch et al. Akademie Verlag. Berlin, p. 175 

\noindent Kronberg, P.P. \& Perry, J.J. 1982, ApJ, 263, 518 

\noindent Kronberg, P.P., Perry, J.J. \& Zukowski, E.L.H. 1992, ApJ, 387, 528 

\noindent Kulsrud, R. 1986, Procc. Joint Varenna-Abastumani Intern. School \& Workshop on Plasma Astrophysics

\noindent Kulsrud, R.M. \& Anderson, S.W. 1992, ApJ, 396, 606

\noindent Lesch, H. \& Chiba, M. 1995, AA, 297, 305 

\noindent Nelson, A.H. 1988, MNRAS, 233, 115

\noindent Owen, F.N., Eilek, J.A., Keel, W.C., 1990, ApJ 362, 449 

\noindent Perley, R.A., Taylor, G.B., 1991, AJ 101, 1623 

\noindent Raphely, Y. \& Gruber, D.E. 1988, ApJ, 333, 133 

\noindent Reuter, H.P., Klein, U., Lesch, H., Wielebinski, R. \& Kronberg, P.P. 1982, AA, 256, 10

\noindent Rozas, M. 1996, Ph. D. Thesis. Univ. La Laguna 

\noindent Ruzmaikin, A.A., Shukurov, A.M. \& Sokoloff, D.D. 1988. Magnetic fields in galaxies (Dordrecht: Kluwer Ac. Pub.) 

\noindent Ruzmaikin, A.A., Sokoloff, D.D. \& Shukurov, A.M. 1989, MNRAS, 241, 1 

\noindent Shapiro, P.R. \& Field, G.B. 1976, ApJ, 205, 762 

\noindent Sofue, Y., Wakamatsu, K. \& Malin, D.F. 1994, Astron. J., 108, 2102 

\noindent Sokoloff, D.D. \& Shukurov, A.M. 1990, Nature, 347, 51 

\noindent Taylor, G.B., Perley, R.A., Inoue, M. et al, 1990, ApJ 360, 41 

\noindent Taylor, G. B., \& Perley, R. A. 1993, ApJ 416, 554 

\noindent Vallee, J.P., MacLeod, J.M. \& Broten, N.W. 1986, AA, 156, 386 

\noindent Watson, A. M., Perry, J. J. 1991, MNRAS 248, 58 

\noindent Wielebinski, R. \& Krause, F. 1993, AA Rev., 4, 449 

\noindent Wielebinski, R. 1993, in The Cosmic Dynamo, ed. F. Krause et al. (Kluwer), 271

\noindent Wolfe, A.M. 1988, in QSO Absorption Lines: Probing the Universe. Ed. by C.V. Blader et at, Cambridge University Press, 297 

\noindent Wolfe, A.M., Lanzetta, K.M. \& Oren, A.L. 1992, ApJ, 387, 17 

\end{document}